\documentclass[aps,prd,reprint,superscriptaddress,nofootinbib]{revtex4-1}

\usepackage{verbatim}
\usepackage{amssymb}
\usepackage{amsmath}
\usepackage{amsfonts}
\usepackage{bm}
\usepackage{hyperref}
\usepackage{graphicx}
\usepackage{tensor}
\usepackage{mathrsfs}
\usepackage{color}
\usepackage{scrextend}
\usepackage{lipsum}

\hypersetup{colorlinks=true,
allcolors=blue
}

\usepackage{enumitem}

\newcommand{\nnm}{\nonumber}
\newcommand{\doe}{\partial}
\newcommand{\be}{\begin{equation}}
\newcommand{\ee}{\end{equation}}
\newcommand{\bea}{\begin{eqnarray}}
\newcommand{\eea}{\end{eqnarray}}
\newcommand{\bdm}{\begin{displaymath}}
\newcommand{\edm}{\end{displaymath}}
\newcommand{\bse}{\begin{subequations}}
\newcommand{\ese}{\end{subequations}}

\newcommand{\mr}{\mathrm}
\newcommand{\tr}{\textrm}
\newcommand{\mc}{\mathcal}

\newcommand{\bs}{\boldsymbol}

\newcommand{\ms}{\mathsf}

\begin{document}

\title{Gravitomagnetic tidal effects in gravitational waves \\from neutron star binaries}
\date{\today}
\author{Batoul Banihashemi}
\email{baniha@umd.edu}
\affiliation{Department of Physics, University of Maryland, College Park, MD 20742, USA}
\author{Justin Vines}
\email{justin.vines@aei.mpg.de}
\affiliation{Albert Einstein Institute, Max Planck Institute for Gravitational Physics, D-14476 Potsdam-Golm, Germany}

\date{\today}

\begin{abstract}

Gravitational waves emitted by coalescing binary systems containing neutron stars (or other compact objects) carry signatures of the stars' internal equation of state, notably, through the influence of tidal deformations during the binary's inspiral stage.  While the leading-order tidal effects for post-Newtonian binaries of compact bodies in general relativity are due to the bodies' mass-quadrupole moments induced by gravitoelectric tidal fields, we consider here the leading effects due to current-quadrupole moments induced by gravitomagnetic tidal fields.  We employ an effective action approach to determine the near-zone gravitational field and the conservative orbital dynamics, initially allowing for arbitrary (not just tidally induced) current-quadrupoles; our approach significantly reduces the complexity of the calculation compared to previous derivations of the conservative dynamics for arbitrary multipoles.
We finally compute the leading contributions from gravitomagnetic tides to the phase and (for the first time) the mode amplitudes of the gravitational waves from a quasi-circular binary inspiral, given in terms of the bodies' quadrupolar gravitomagnetic tidal Love numbers (tidal linear response coefficients in an adiabatic approximation).    In the phase, gravitomagnetic tides are suppressed by one post-Newtonian order relative to gravitoelectric ones, but this is not always the case for the mode amplitudes.  In the $(\ell,|m|)=(2,1)$ and $(3,2)$ modes for example, they appear at the same leading orders.
\end{abstract}

\maketitle

\section{Introduction}

The much anticipated prospect of using gravitational wave (GW) observations to probe the internal structure of neutron stars (NSs) through the effects of tidal interactions in NS binaries \cite{Flanagan:2007ix,Read:2009yp,Lackey:2011vz,Pannarale:2011pk,Damour:2012yf,DelPozzo:2013ala,Wade:2014vqa,Bernuzzi:2014kca,Agathos:2015uaa,Hotokezaka:2016bzh,Favata:2013rwa,Yagi:2013baa,Cullen:2017oaz} has finally been realized with the first detection by LIGO-Virgo of GWs from a binary NS merger, GW170817 \cite{TheLIGOScientific:2017qsa}.  Among the multitude of astrophysical questions that can be addressed using GW170817 together with the coincident trans-spectral electromagnetic observations of the NS merger and its aftermath \cite{GBM:2017lvd}, the measurement of tidal effects is one which relies only on the GW signal.  As pointed out in \cite{Flanagan:2007ix}, the internal structure of each NS, in particular its equation of state (EoS) \cite{Lattimer:2006xb}, influences the inspiral-stage GW signal primarily via a single parameter $\lambda$ known as the star's (quadrupolar gravitoelectric) tidal deformability/polarizability, or tidal Love number (TLN), which measures the star's leading-order tidal deformation, its induced mass-quadrupole moment, in response to its companion's gravitational field.  The LIGO-Virgo collaboration has used GW170817 to place an upper bound on the binary's effective TLN (a certain mass-weighted sum of the TLNs of the two NSs) \cite{TheLIGOScientific:2017qsa}, and this rather tangibly translates into an upper bound of around 13 km for the NS radii (independently of the uncertainty in the mass ratio and the spins) \cite{Raithel:2018ncd}, already ruling out some of the stiffer candidate EoSs.

The properties of the binary NS were inferred in \cite{TheLIGOScientific:2017qsa} by matching the GW signal to post-Newtonian (PN) \cite{Blanchet:2013haa,poisson2014gravity} frequency-domain waveform models \cite{Sathyaprakash:1991mt} for binary inspirals in general relativity (GR).  The physics included in such models naturally splits into various contributions to the inspiral dynamics associated with the stars' various \emph{multipole moments}.  

Contributions which are independent of the nature of the bodies comprising the binary, and which contribute at the lowest orders in the PN approximation, arise from the leading terms in the bodies' multipole expansions: (i) the ``point-mass'' contributions, depending only on the bodies' masses (their \emph{mass-monopoles}), and (ii) contributions which depend linearly on the bodies' spins (their angular momenta, or \emph{current-dipoles}).  The inspiral dynamics at the point-mass and linear-in-spin levels is the same for a binary NS as for a binary black hole, or any compact binary in GR,\footnote{The universality (body-independence) of the dynamics holds for the contributions which are separately linear in each body's spin, including not only the ``spin-orbit'' terms which are linear in one spin but also the $S_1$-$S_2$ terms, which should be classified here as part of the ``linear-in-spin" dynamics.} and it has been calculated within the PN framework \cite{Arun:2008kb,Blanchet:2013haa,Mishra:2016whh}, as well as within the effective-one-body \cite{Buonanno:1998gg,Buonanno:2000ef,Damour:2008qf,Damour:2009kr,Barausse:2009xi,Nagar:2017jdw} and phenomenological \cite{Ajith:2007qp,Ajith:2009bn,Hannam:2013oca,Husa:2015iqa,Khan:2015jqa,Smith:2016qas} approaches which combine PN (and other analytical) information with data from numerical relativity simulations of binary black holes.  

A body's internal structure influences the inspiral dynamics through contributions from its higher multipole moments, starting with the \emph{mass-quadrupole} and the \emph{current-quadrupole}, on through the infinite set of \emph{mass-multipoles} which couple to the \emph{gravitoelectric} tidal fields, and the infinite set of \emph{current-multipoles} which couple to the \emph{gravitomagnetic} tidal fields \cite{Damour:1990pi,Damour:1991yw,Racine:2004hs,poisson2014gravity}.

The most important higher multipole, the body's mass-quadrupole, represented by a symmetric trace-free (STF) spatial tensor $M_{ij}$ \cite{poisson2014gravity}, is determined in an adiabatic approximation\footnote{The adiabatic approximation assumes that the body's internal dynamical time scales are much less than the orbital timescale, which is a natural leading assumption for binaries of compact objects in the PN regime.} by two leading effects: (i) an intrinsic spin-induced (oblate) deformation scaling as the square of the spin $S_i$ \cite{Poisson:1997ha}, superposed with (ii) a tidally induced (prolate) deformation proportional to the quadrupolar gravitoelectric tidal field $\mc E_{ij}$ due to the companion,\footnote{At Newtonian (0PN) order, the (gravitoelectric) quadrupolar tidal field is given in our conventions by $\mc E_{ij}=-\doe_i\doe_jU$, where $U$ is the Newtonian potential due to the companion, with $U=Gm'/r$ for a point-mass companion $m'$.  The angle brackets $<\ldots>$ in (\ref{quad_intro}) denote STF projection of the enclosed indices.}
\be\label{quad_intro}
M_{ij}=-\frac{\kappa}{mc^2}S_{< i}S_{j>}-\lambda \mc E_{ij}+\ldots
\ee
The coefficient $\kappa$, determining the spin-squared quadrupole, and the coefficient $\lambda$, which is the quadrupolar gravitoelectric TLN, depend on the nature of the body.  For a black hole, $\kappa=1$ and $\lambda=0$ \cite{Hansen:1974zz,Poisson:1997ha,Porto:Rothstein:2008:2, Taylor:2008xy,Damour:2009vw,Kol:2011vg,Pani:2015hfa}, while for a NS, $\kappa>1$ and $\lambda>0$ are determined by the NS EoS (and the NS's mass $m$) \cite{Laarakkers:1997hb,Hinderer:2007mb,Hinderer:2009ca,Binnington:2009bb,Damour:2009vw,Postnikov:2010yn}.  For the spin-squared mass-quadrupole $\propto\kappa$, the inspiral dynamics has been computed at the leading (2PN) and next-to-leading (3PN) orders \cite{Poisson:1997ha,Arun:2008kb,Mishra:2016whh,Mikoczi:2005dn,Bohe:2015ana}.\footnote{The conservative dynamics has been computed at 4PN order \cite{Levi:Steinhoff:2015:3}.}
For the adiabatic tidal mass-quadrupole $\propto\lambda$, the inspiral dynamics has been computed at the leading (formally 5PN) and next-to-leading (6PN) orders \cite{Flanagan:2007ix,Damour:2009wj,Vines:2010ca,Vines:2011ud,Bini:2012gu,Damour:2012yf,Steinhoff:2016rfi}.\footnote{The conservative dynamics has been computed at 7PN order \cite{Bini:2012gu}.}  Precisely these structure-dependent contributions were added to the point-mass and linear-in-spin dynamics (through 3.5PN order) in the frequency-domain PN waveforms used to infer the properties of GW170817 in \cite{TheLIGOScientific:2017qsa}.\footnote{Going beyond the PN framework, tidal effects have been included in effective-one-body models \cite{Damour:2009wj,Baiotti:2010xh,Bini:2012gu,Bini:2012gu,Bini:2014zxa,Bernuzzi:2014owa,Hotokezaka:2015xka,Hinderer:2016eia,Steinhoff:2016rfi}, which have been compared to numerical simulations of NS binaries \cite{Baiotti:2010xh,Baiotti:2011am,Bernuzzi:2012ci,Hotokezaka:2013mm,Bernuzzi:2014owa,Hotokezaka:2015xka,Hinderer:2016eia,Dietrich:2017feu}, including dynamical (non-adiabatic) tidal effects in \cite{Hinderer:2016eia,Steinhoff:2016rfi,Dietrich:2017feu}, and in other phenomenological or semi-analytical approaches based on numerical simulations and PN information, e.g.\ \cite{Ferrari:2011as,Dietrich:2017aum,Dietrich:2018uni,Kawaguchi:2018gvj}.}  As emphasized in \cite{Wade:2014vqa,Favata:2013rwa,Yagi:2013baa,Cullen:2017oaz,Harry:2018hke}, the measurement of tidal effects (or of masses and spins) is sensitive to the accuracy with which the point-mass, spin, and tidal dynamics are described; exclusion of higher-order terms can significantly bias the parameter estimation in some circumstances.  

Apart from higher-PN-order corrections to the point-mass dynamics, the spin terms, and the mass-quadrupole $M_{ij}$ terms, we encounter at higher orders the contributions from the bodies' higher (STF) multipoles:  the mass-octupole $M_{ijk}$, mass-hexadecapole $M_{ijkl}$, etc., and the current-quadrupole $S_{ij}$, current-octupole $S_{ijk}$, etc.  For each multipole, in the adiabatic approximation, the leading contributions arise again from spin-induced and tidal deformations, according to
\begin{alignat}{7}
M_{ij}&\;=\;\bar\kappa_2\, S_{< i}S_{j>}&\;+\,\,&\bar\lambda_2\, \mc E_{ij}&&+\ldots
\\\nnm
S_{ij}&\;=\;&&\bar\sigma_2\,\mc B_{ij}&&+\ldots
\\\nnm
M_{ijk}&\;=\;&&\bar\lambda_3\,\mc E_{ijk}&&+\ldots
\\\nnm
S_{ijk}&\;=\;\bar\kappa_3\, S_{< i}S_jS_{k>}&\;+\,\,&\bar\sigma_3\,\mc B_{ijk}
&&+\ldots
\\\nnm
M_{ijkl}&\;=\;\bar\kappa_4\, S_{< i}S_jS_kS_{l>}&\;+\,\,&\bar\lambda_4\,\mc E_{ijkl}&&+\ldots
\\\nnm
S_{ijkl}&\;=\;&&\bar\sigma_4\,\mc B_{ijkl}&&+\ldots
\end{alignat}
etc., with spin-induced $\bar\kappa_\ell$ terms only for the even-$\ell$ mass-multipoles and the odd-$\ell$ current-multipoles [entering at $\ell$PN and \mbox{$(\ell+\frac{1}{2})$PN} orders respectively, where the multipole order $\ell\ge2$ is the number of indices $ij\ldots$], with tidal $\bar\lambda_\ell$ terms for all the mass-multipoles proportional to the gravitoelectric tidal tensors $\mc E_{ij\ldots}$ [entering at $(2\ell+1)$PN order], and with tidal $\bar\sigma_\ell$ terms for all the current-multipoles proportional to the gravitomagnetic tidal tensors $\mc B_{ij\ldots}$ [entering at $(2\ell+2)$PN order]. Further contributions to the multipoles, still within the adiabatic approximation, will arise e.g.\ from spin-tidal couplings (e.g.\ adding a term $\propto\mc B_{ijk}S^k$ to $M_{ij}$), proportional to the rotational-tidal Love numbers \cite{Pani:2015hfa,Pani:2015nua,Landry:2017piv,Gagnon-Bischoff:2017tnz,Abdelsalhin:2018reg,Landry:2018bil} [entering at 6.5PN order or higher].

The leading gravitomagnetic tidal term, the $\bar\sigma_2$ term in the current-quadrupole $S_{ij}$, which (along with other more general results concerning $S_{ij}$ effects in binaries) is the subject of this paper, enters the inspiral dynamics at 6PN order, the same order as the next-to-leading corrections to the leading quadrupolar gravitoelectric tidal term $\propto\bar\lambda_2$.  These are the only tidal terms contributing at 6PN order or less in the adiabatic approximation.

The process of determining how adiabatic tidal deformations influence a binary inspiral and its GW emissions---assuming some EoS for each body, assuming GR, and treating the binary dynamics in the PN approximation\footnote{While we assume, for the validity of the PN approximation, that the binary (with masses $\sim m$, orbital radius $r$, and speed $v$) satisfies the weak-field, slow-motion assumption $Gm/rc^2\sim v^2/c^2\ll1$, this does not mean that the gravitational field must be weak everywhere in the system.  Each body could have strong internal gravity (as is the case for NSs and black holes), as long as (roughly speaking) the field of one body is weak at the location of the other; see \cite{Racine:2004hs}.  The calculations of TLNs for NSs and black holes are strong-field (not PN) calculations, the outputs of which are fed into the PN treatment of the binary inspiral.}---consists of three main steps:
\begin{itemize}
\item \emph{Calculate TLNs}:  given a star's EoS and its mass, solve (in full GR) for the equilibrium stellar structure and then for linear perturbations induced by (stationary) asymptotic tidal fields, determining the TLNs (which enter as parameters in the following two steps) as the proportionality constants between the induced multipoles and the perturbing tidal fields.

\item \emph{Conservative dynamics}:  determine solutions to the PN field equations in the near zone along with conservative orbital equations of motion for the binary (and/or an action principle from which they derive).  This step could be done from the beginning for the special case of adiabatic tidally induced multipoles, or done for arbitrary multipoles and then specialized.

\item \emph{Radiative dynamics}:  determine the GWs sourced by the conservative motion, and the resultant radiation reaction which drives the inspiral (e.g.\ by calculating the GW energy flux and enforcing energy-balance).  A crucial output is the tidal contribution to the phase of the frequency-domain PN waveform; one can additionally compute tidal contributions to the amplitudes of the tensor-spherical-harmonic modes of the waveform.

\end{itemize}

In the following we review work on each of these steps, (mostly) pertaining to the adiabatic quadrupolar gravitoelectric tidal effects with $M_{ij}=\bar\lambda_2\mc E_{ij}$ and the adiabatic quadrupolar gravitomagnetic tidal effects with $S_{ij}=\bar\sigma_2\mc B_{ij}$.  We also describe and put into context the original work of this paper, which addresses the last two steps for the $S_{ij}$ case.

\subsubsection{Tidal Love numbers}

The quadrupolar gravitoelectric TLN $\lambda$,
\be
\lambda=-\bar\lambda_2=-\frac{M_{ij}}{\mc E_{ij}},
\ee
has been computed for relativistic stars in GR with a variety of EoSs, including polytropes, realistic NS models, and quark star models \cite{Hinderer:2007mb,Hinderer:2009ca,Binnington:2009bb,Damour:2009vw,Postnikov:2010yn}.  
It can be expressed in terms of the stellar radius $\mc R$ and the dimensionless apsidal constant $k_2^\mr{el}$ as
\be\label{lambdak}
\lambda=\frac{2}{3G}k_2^\mr{el}\mc R^5,
\ee
with $k_2^\mr{el}$ in the range $\sim$ 0.05 -- 0.15 for realistic hadronic EoSs \cite{Hinderer:2007mb,Hinderer:2009ca,Postnikov:2010yn}.\footnote{Our definition of $\lambda$ matches the $\lambda$'s defined in \cite{Hinderer:2007mb,Hinderer:2009ca,Vines:2010ca,Vines:2011ud} and is the same as the $\mu^{(2)}$ in \cite{Bini:2012gu}.  Our $k_2^\mr{el}$ matches that in \cite{Gagnon-Bischoff:2017tnz}.}

The quadrupolar gravitomagnetic TLN $\sigma$,
\be
\sigma=\frac{\bar\sigma_2}{2c^2}=\frac{S_{ij}}{2c^2\mc B_{ij}},
\ee
has been computed for various EoSs (and with different assumptions on the fluid state) e.g.\ in \cite{Binnington:2009bb,Damour:2009vw,Landry:2015cva,Pani:2015hfa,Pani:2015nua,Landry:2017piv,Gagnon-Bischoff:2017tnz}.  It can be expressed as
\be\label{sigmaR}
\sigma=\frac{1}{2c^2}k_2^\mr{mag} m\mc R^4,
\ee
where $m$ is the stellar mass and $k_2^\mr{mag}$ is dimensionless.\footnote{Our $\sigma$ matches the $\sigma^{(2)}$ in \cite{Bini:2012gu}.  Our $k_2^\mr{mag}$ matches that in \cite{Gagnon-Bischoff:2017tnz}.   Our conventions for the normalizations of $\mc E_{ij}$ and $\mc B_{ij}$ are given in (\ref{EBR}) below, and our $M_{ij}$ and $S_{ij}$ match the Blanchet-Damour multipoles \cite{Blanchet:1989ki}; see Footnote \ref{foot:norm}.}

Following an analysis of gravitomagnetic tidal effects in the context of PN gravity in \cite{Favata:2005da}, $\sigma$ was first computed for fully relativistic stars in \cite{Binnington:2009bb,Damour:2009vw}.  The latter references took the stellar fluid to be in a state of strict hydrostatic equilibrium, the ``static state,'' and obtained positive values for $\sigma$. It was subsequently argued in \cite{Landry:2015cva} that the fluid should be in a non-static ``irrotational state,'' with nonzero tidally induced currents but with the fluid having vanishing vorticity, and calculations with this assumption yielded negative values for $\sigma$.  

An important development in the study of tidal deformations was the discovery in \cite{Yagi:2013bca} of approximately universal, EoS-independent relations between the TLN $\lambda$ and the spin-induced quadrupole coefficient $\kappa$ (and between each of those and the star's moment of inertia $I$).  The universality has been linked to the emergence of self-similarity of the stars' isodensity contours \cite{Yagi:2014qua}.  Such universal relations have been analyzed for higher (gravitoelectric and -magnetic) multipoles \cite{Yagi:2013sva,Yagi:2014bxa,Pappas:2013naa,Chatziioannou:2014tha,Delsate:2015wia}, for rotational-tidal Love numbers \cite{Pani:2015hfa,Pani:2015nua,Landry:2017piv,Gagnon-Bischoff:2017tnz}, for the case of rapid rotation \cite{Doneva:2013rha,Chakrabarti:2013tca}, for dynamical configurations \cite{Maselli:2013mva}, for magnetized NSs \cite{Haskell:2013vha}, in alternative gravity theories \cite{Yagi:2013bca,Sham:2013cya,Doneva:2014faa,Pani:2014jra,Kleihaus:2014lba,Doneva:2017jop,Mariji:2017buj}, for extreme EoSs \cite{Silva:2017uov}, and for use in parameter estimation with binary inspiral GWs \cite{Yagi:2013sva,Agathos:2015uaa,Yagi:2015pkc,Chatziioannou:2018vzf}; see \cite{Yagi:2016bkt} for a review.  Regarding some of the most recent calculations of particularly $\lambda$ and $\sigma$ for realistic NS EoSs, we note that calculations using both the ``static'' \cite{Pani:2015nua} and ``irrotational'' \cite{Gagnon-Bischoff:2017tnz} fluid assumptions show universality for $\lambda$ and $\sigma$ at a $\sim$1\% level (whereas for the rotational-TLNs, \cite{Pani:2015nua} finds significant deviations from universality while \cite{Gagnon-Bischoff:2017tnz} finds universality at a $\sim$2.5\% level).

\subsubsection{Conservative orbital dynamics}

To determine the effects of the tidal deformations (or more generally, of any nonzero higher multipoles) on the binary dynamics according to GR, one must construct appropriately parametrized solutions to the Einstein field equations in the binary's near zone.  This problem has been treated, in principle to all multipolar orders, at the level of the 1PN field equations (which is sufficient to treat the mass-multipoles at leading and next-to-leading PN orders, and the current multipoles at leading orders) in a series of works by Damour, Soffel and Xu and Racine and Flanagan \cite{Damour:1990pi,Damour:1991yw,Damour:1992qi,Racine:2004hs,Damour:1993zn,Racine:2004xg} which developed a general formalism for 1PN celestial mechanics with arbitrarily structured bodies; we will refer to this as the DSX-RF formalism.  It provides near-zone solutions to the field equations along with translational and rotational equations of motion for a system of $n$ bodies with arbitrary multipoles, specifying the system's conservative dynamics.  In \cite{Vines:2010ca}, the DSX-RF formalism was directly applied to compute the conservative dynamics for a binary of bodies having arbitrary mass-quadrupoles $M_{ij}(t)$, through next-to-leading order in the $M_{ij}$ terms (to linear order in $M_{ij}$'s), finding an action principle governing the orbital dynamics \emph{post hoc} by matching to the equations of motion; the results for arbitrary $M_{ij}$ were finally specialized to the case of adiabatic tidally induced $M_{ij}$.   A considerably simpler route to an action for the conservative dynamics, specializing from the beginning to the case of adiabatic tides, was provided in \cite{Damour:2009wj,Bini:2012gu} via a covariant effective action treatment; results for the $M_{ij}$ case through next-to-leading order were first presented in \cite{Damour:2009wj} and were derived in \cite{Bini:2012gu} along with next-to-next-to-leading $M_{ij}$ and leading and next-to-leading $S_{ij}$ and $M_{ijk}$ results.  Very recently, the leading conservative dynamics contributions due to adiabatic rotational-tidal effects have been computed in \cite{Abdelsalhin:2018reg,Landry:2018bil}; both works employ the orbital equations of motion derived from the DSX-RF formalism.

In this paper, in Sec.~\ref{sec:cons}, we first show how one can efficiently compute the leading-order contributions to a binary's conservative orbital dynamics from \emph{arbitrary} current-quadrupoles $S_{ij}(t)$ via a simple covariant effective action treatment, following e.g.\ the treatment of arbitrary $M_{ij}(t)$ in \cite{Steinhoff:2016rfi}.\footnote{It was shown in \cite{Steinhoff:2016rfi} that the conservative dynamics with arbitrary $M^{ij}(t)$ is defined in the effective field theory approach by the formal body action appearing below in our Eq.~(\ref{SEB}) (without the $S^{\mu\nu}$ term, and with the modification discussed in Footnote \ref{footpsi}), and that an equivalent form of the next-to-leading-order arbitrary-$M^{ij}(t)$ binary action from \cite{Vines:2010ca} can be reproduced by taking results derived for the spin-squared quadrupole case and making the replacement $\bar\kappa_2 S_{<i}S_{j>}\to M_{ij}$.} We then obtain the specialized results for the adiabatic case, $S_{ij}\propto\mc B_{ij}$, with an appropriate ansatz for an action for the body's (adiabatic) internal dynamics, reproducing (the leading-order part of) the adiabatic tidal $S_{ij}$ results in \cite{Bini:2012gu}.

We emphasize that the effective action approach with arbitrary multipoles, as implemented here for current quadrupoles, fully reproduces the near-zone dynamics derived from the DSX-RF formalism, but with much less computational effort (as was the case for the effective action approach specialized to adiabatic tidal multipoles in \cite{Bini:2012gu}).  The DSX-RF formalism works at the level of the field equations and orbital equations of motion and employs multipole coordinate systems (one global one and one adapted to each body) with intricate transformations between them; (relatively compact) action principles for the orbital dynamics have been deduced \emph{post hoc} by matching to the (relatively lengthy) equations of motion \cite{Vines:2010ca,Abdelsalhin:2018reg}.  Our approach here employs a single global coordinate system and arrives directly at an action principle encoding both the field equations and the orbital dynamics, the form of which is entirely fixed (at leading order) by the requirement of general covariance of the effective action.

\subsubsection{Gravitational radiation}

The GW emissions and radiation reaction can be determined by appropriately matching a near-zone PN solution to a far-zone multipolar post-Minkowskian solution, as reviewed in \cite{Blanchet:2013haa}.  The matching results in expressions for the far-zone GW field given in terms of the binary system's radiative multipole moments, which are related in certain ways to the system multipole moments which are encoded in the near-zone PN solution, as reviewed in Sec.~3 of \cite{Blanchet:2013haa}. 
From the radiative multipoles, one can compute the GW energy flux, and for (nonspinning or aligned-spin) quasi-circular binaries, balancing the flux against the energy loss from the near-zone dynamics determines the rate of inspiral and allows a computation of the phase of the frequency-domain GW signal, as in Sec.~9 of \cite{Blanchet:2013haa}.  In addition to the phase, one can compute the amplitudes of the tensor-spherical-harmonic modes of the far-zone field directly from the radiative multipoles.
All of this formalism, which has been developed and applied at high orders for point-mass and spin effects, is directly applicable to binaries with bodies having arbitrary higher multipoles.

The first analytic calculations of tidal effects in GW signals were presented in \cite{Flanagan:2007ix}, which derived the contributions to the phase from the leading adiabatic $M_{ij}$ tides, at leading PN (Newtonian) order (which count as 5PN-order contributions in terms of their scaling with frequency).  The next-to-leading (6PN) adiabatic tidal $M_{ij}$ corrections to the phase were computed in \cite{Vines:2011ud}, and the corresponding leading and next-to-leading terms in the mode amplitudes were computed in \cite{Damour:2012yf}.  The leading $(2\ell+1)$PN-order phase contributions from gravitoelectric tides, with adiabatic tidal $M_{i_1\ldots i_\ell}$ for arbitrary $\ell$, were derived in \cite{Yagi:2013sva}.  The 6PN phase contribution from the leading gravitomagnetic tides, with adiabatic tidal current-quadupoles $S_{ij}$, was first derived completely in \cite{PhysRevD.96.129904}, correcting omissions in an initial calculation in \cite{Yagi:2013sva}, and later being confirmed in \cite{Abdelsalhin:2018reg}.  Very recently, the 6.5PN phase contributions due to rotational-tidal couplings have been computed in \cite{Abdelsalhin:2018reg,Landry:2018bil}.

In this paper, in Sec.~\ref{sec:rad}, we reproduce the recent results of \cite{PhysRevD.96.129904,Abdelsalhin:2018reg} for the 6PN gravitomagnetic tidal contributions to the GW flux and phase, and we present for the first time the corresponding contributions to the GW mode amplitudes, $\propto\sigma$.  We compute the $(\ell,m)$ mode amplitudes for $\ell=2$ and 3 including all adiabatic tidal effects through relative 1PN order.  This includes new contributions from quadrupolar gravitoelectric tides, $\propto\lambda$, given here for the first time, in particular the next-to-leading $\lambda$ terms in the $(3,3)$ and $(3,1)$ modes and the leading $\lambda$ term in the $(3,2)$ mode.  We point out that $\lambda$ and $\sigma$ terms appear at the same leading PN orders in the $(3,2)$ and $(2,1)$ modes, in contrast to the modes with $\ell+m$ = even (and the phase) where $\sigma$ terms are suppressed by one PN order relative to $\lambda$ terms.  While the tidal ($\lambda$ and $\sigma$) contributions to the $(2,1)$ mode (and more generally the odd-$m$ modes) vanish for equal masses and equal Love numbers, they do not vanish in that case for the $(3,2)$ mode (and more generally the even-$m$ modes).

\section{Conservative dynamics}\label{sec:cons}

\subsection{Covariant effective action for monopole-quadrupole bodies}\label{sec:generalS}

It is well known that the gravitational dynamics, in GR, of a system of $n$ (monopolar) point-masses $m_\ms{a}$, with $\ms a=1,\ldots,n$, having arbitrarily parametrized worldlines $x^\mu=z_\ms{a}^\mu(s_\ms{a})$ with tangents $\dot z^\mu_\ms{a}=dz_\ms{a}/ds_\ms a$ in a spacetime with metric $g_{\mu\nu}(x)$, is formally defined by the action
\be\label{pmaction}
\mc S[z_\ms{a},g]=-\sum_\ms{a}m_\ms{a}c^2\int d\tau_\ms{a}+\mc S_g[g],
\ee
where $\mc S_g$ is the Einstein-Hilbert action and
\be
d\tau_\ms{a}=ds_\ms{a}\sqrt{-g_{\mu\nu}(z_\ms{a})\dot z_\ms{a}^\mu\dot z_\ms{a}^\nu},
\ee
using the $(-,+,+,+)$ signature.  Varying the action with respect to one of the worldlines yields the geodesic equation, and varying with respect to the metric yields Einstein's equation with a distributional stress-energy tensor along the worldlines.  These coupled equations can be solved consistently e.g.\ in the PN approximation, given a way to properly regularize the infinite self-field contributions (which, for our purposes below, amounts to simply dropping all such infinite terms).

In order to describe, instead of a point-mass, an extended body with internal structure and dynamics, but which is still localized on a scale small compared to the (external) spacetime's radius of curvature, it is natural to maintain the description in terms of a worldline $x^\mu=z^\mu(s)$, but now with additional degrees of freedom $\psi(s)$ defined along the worldline.  In an effective action approach, the way in which these degrees of freedom can couple to the external gravitational field is tightly constrained by general covariance \cite{Blanchet:2013haa,Porto:2016pyg,Levi:Steinhoff:2015:1,Steinhoff:2016rfi}.  An effective Lagrangian can depend on the worldline point $z$ only through the metric and invariant curvature tensors (the Riemann tensor and its covariant derivatives) evaluated at $z$,\footnote{\label{footpsi}The Lagrangian could also depend on the worldline point and metric through covariant parameter derivatives $D\psi/ds$ of the worldline fields $\psi$, as is the case for actions encoding spin effects \cite{Blanchet:2013haa,Porto:2016pyg,Levi:Steinhoff:2015:1} and for the action for dynamical mass-quadrupole tides in \cite{Steinhoff:2016rfi}, both of which require (only) the generalization $\mc L^\mr{int}[\psi]\to\mc L^\mr{int}[\psi,g,z]$ in (\ref{SEB}), with the metric- and $z$-dependence coming only from $D\psi/ds$.  Here, we do not treat spin (current-dipole) effects, as they are entirely decoupled from the \emph{leading-order} quadrupole effects we analyze here, and our ansatz (\ref{SEB}) is sufficient for this analysis.  Note for example that the metric-dependence via $D\psi/ds$ in $\mc L^\mr{int}$ seen in the first term of Eq.~(1.4) of \cite{Steinhoff:2016rfi} affects the orbital dynamics only at next-to-leading order, producing to the ``frame-dragging Lagrangian'' in Eq.~(3.12) of \cite{Steinhoff:2016rfi}; those contributions are related to the fact that next-to-leading order mass-quadrupole effects cannot be disentangled from spin effects, as discussed in \cite{Vines:2010ca}.} and the leading terms will thus be those linear in the (regularized) Riemann/Weyl tensor $R_{\mu\nu\rho\sigma}$ (of the exterior metric, which we assume is a vacuum solution, so the Riemann tensor equals the Weyl tensor).\footnote{The fact that we can cleanly split $R_{\mu\nu\rho\lambda}$ (or any functional of the metric), at the location of some localized body, into an internal part (locally generated by the body) and an external part (due to the rest of the universe) corresponding to a vacuum solution, for our purposes below, is ensured by the linearity of the 1PN field equations (\ref{UFE})--(\ref{UiFE}).} It is natural to split $R_{\mu\nu\rho\sigma}$ into its electric (even-parity) and magnetic (odd-parity) parts with respect to the worldline's normalized 4-velocity $u^\mu=dz^\mu/d\tau=\dot z^\mu/\sqrt{-\dot z^2}$ according to
 \begin{equation}\label{EBR}
  \mc{E}_{\mu\nu}=R_{\mu\rho\nu\sigma}u^{\rho}u^{\sigma}  , \qquad \mc{B}_{\mu\nu}=R^*_{\mu\rho\nu\sigma}u^{\rho}u^{\sigma},
\end{equation}
where $R^*_{\mu\rho\nu\sigma}=\frac{1}{2}\epsilon _{\nu\sigma}^{\quad \alpha\beta}R_{\mu\rho\alpha\beta}$ is the dual of $R_{\mu\rho\nu\sigma}$, both evaluated at the worldline point $z$.  Both $\mc{E}_{\mu\nu}$ and $\mc{B}_{\mu\nu}$ are STF and orthogonal to $u^{\mu}$. The full information of $R_{\mu\nu\rho\sigma}$ can be reconstructed from $\mc{E}_{\mu\nu}$, $\mc{B}_{\mu\nu}$, and $u^{\mu}$, e.g.\ as in Eq.~(5.12) of \cite{Vines:2016unv}.   We can then write a general ansatz for the action for our extended body, to linear order in $R_{\mu\nu\rho\sigma}$, by replacing $-mc^2\int d\tau$ in (\ref{pmaction}) with
\begin{alignat}{3}\label{SEB}
&\mc S_\mr{body}[z,\psi,g]=
\\\nnm
&\int d\tau\left(-mc^2-\frac{1}{2}M^{\mu\nu}[\psi]\mc E_{\mu\nu}+\frac{2}{3}S^{\mu\nu}[\psi]\mc B_{\mu\nu}+\mc L^\mr{int}[\psi]\right),
\end{alignat}
where the coefficients $M^{\mu\nu}$ and $S^{\mu\nu}$, which we will identify with the mass- and current-quadrupole moments,\footnote{The factors of -1/2 and 2/3 in (\ref{SEB}) are simply matters of convention for the normalization of the multipoles.  See Footnote \ref{foot:norm}.} and the internal Lagrangian $\mc L^\mr{int}$ are some functionals of only the internal degrees of freedom $\psi$.  We take $M^{\mu\nu}$ and $S^{\mu\nu}$ to be STF and orthogonal to $u^\mu$, as only these components contribute in (\ref{SEB}).  

The body action (\ref{SEB}), added to actions for other bodies and the gravity action $S_g$, determines the body's orbital dynamics regardless of the details of the functionals $M^{\mu\nu}$, $S^{\mu\nu}$, and $\mc L^\mr{int}$; varying the total action with respect to the worldlines and the metric yields orbital equations of motion and field equations in which $M^{\mu\nu}(s)=M^{\mu\nu}[\psi(s)]$ and $S^{\mu\nu}[\psi(s)]$ (and their $s$-derivatives) enter as some thus far arbitrary functions.  A specification of the internal degrees of freedom $\psi$ and of the functionals $\mc L^\mr{int}$, $M^{\mu\nu}$, and $S^{\mu\nu}$ would require some model for the body's internal structure, but we can proceed with an analysis of the orbital dynamics while leaving these details unspecified.

Here we are interested in the effects linear in the current-quadrupoles $S^{\mu\nu}_\ms{a}$, for a binary, $\ms a=1,2$.  Since we would drop $S_1^{\mu\nu}$-$S_2^{\mu\nu}$ terms, we need only consider one of the bodies, say body 1, to have a nonzero quadrupole $S_1^{\mu\nu}$, and we can restore the $S_2^{\mu\nu}$ terms at the end of the calculation by interchanging the bodies' identities.  We thus consider the following action for our binary,
\be\label{binaryaction}
\mc S[z_1,\psi_1,z_2,g]=\mc S_1[z_1,\psi_1,g]+\mc S_2[z_2,g]+\mc S_g[g],
\ee
where we take body 2 to be simply a point mass,
\be
\mc S_2=-m_2c^2\int d\tau_2,
\ee
and the body-1 term is
\be
\mc S_1=\int d\tau_1\left(-m_1c^2+\frac{2}{3}S_1^{\mu\nu}\mc B^1_{\mu\nu}+\mc L_1^\mr{int}\right),
\ee
where $\mc B^1_{\mu\nu}=u_1^\rho u_1^\sigma R^*_{\mu\rho\nu\sigma}(z_1)$, and where the current-quadrupole $S_1^{\mu\nu}[\psi_1]$ and the internal Lagrangian $\mc L_1^\mr{int}[\psi_1]$ are some unspecified functionals of body 1's internal degrees of freedom $\psi_1$.

\subsection{Leading-order post-Newtonian approximation}

We now specialize to the PN approximation and assume that the metric $ds^2=g_{\mu\nu}dx^\mu dx^\nu$ can be written in the following standard 1PN form \cite{Blanchet:2013haa,poisson2014gravity}, through $O(c^{-2})$ relative to the Newtonian metric, in (spatially conformally flat) coordinates $x^{\mu}=(t,x^i)=(t,\bs x)$, (note $x^0=t$ in our conventions; $x^0\ne ct$)
\begin{alignat}{3}  \label{1PNmet}
  ds^2&= - \Big( c^2 -2U+\frac{2U^2}{c^2} \, \Big) \, dt^2 - \frac{8}{c^2} U_i \, dt \, dx^i 
  \nnm\\
  &\quad + \Big( 1+\frac{2U}{c^2} \Big) \delta_{ij} \, dx^i \, dx^j +  O(c^{-4}) ,
\end{alignat}
given in terms of the (gravitoelectric) scalar potential $U(t,\bs{x})$ and the (gravitomagnetic) vector potential $U_i(t,\bs{x})$.
Adopting the harmonic gauge condition,
\be
\partial_{\mu}(\sqrt{-g}g^{\mu\nu})=0\quad\;\Rightarrow\;\quad \dot U+\doe_iU^i=O(c^{-2}),
\ee
Einstein's equation with a stress-energy tensor $T^{\mu\nu}=(T^{00},T^{0i},T^{ij})$ yields the harmonic-gauge 1PN field equations
\begin{align}\label{UFE}
  \bs{\nabla}^2 U & = -4 \pi G \Big( T^{00}+\frac{T^{ii}}{c^2} \Big) + \frac{\ddot{U}}{c^2} +  O(c^{-4}) 
  \\\label{UiFE}
  \bs{\nabla}^2U^i & = -4 \pi G T^{0i} +  O(c^{-2}),
\end{align}
where $\bs\nabla^2=\doe_i\doe^i$ and $T^{ii}\equiv \delta_{ij}T^{ij}$.  Note that we use the Euclidean metric $\delta_{ij}$ to raise and lower the spatial indices $i,j,\ldots$; for components of four-dimensional tensors, where there is possible ambiguity, we leave indices in the position in which the components are taken, or explicitly clarify, as in the case of $T^{ii}= \delta_{ij}T^{ij}$.  In this section, we maintain proper up-down contractions; in the following section, where this becomes less feasible, we switch to the usual convention that up/down placement of indices is irrelevant, and all spatial indices are contracted with $\delta_{ij}$.

The gauge-fixed gravitational action (the Einstein-Hilbert action plus a harmonic gauge-fixing term) for the 1PN metric (\ref{1PNmet}) reads
\begin{equation}\label{Sg1PN}
  S_{g}= \int \frac{dt \, d^3 \bs{x}}{8\pi G} \Big [ -\doe_i U\doe^i U+\frac{\dot U^2}{c^2}+\frac{4}{c^2}\doe_iU_j\doe^iU^j+ O(c^{-4}) \Big ]\:.
\end{equation}
To evaluate the other terms in our binary action (\ref{binaryaction}), we choose both of the worldline parameters to be the coordinate time, $s_1=s_2=t$, so that $\dot z_\ms{a}^\mu=(1,v^i_\ms{a})$ where $v_\ms{a}^i=dz_\ms a^i/dt$, and then the normalized 4-velocities are given by $u^\mu_\ms{a}=\gamma_\ms{a}(1,v_\ms{a}^i)$ with
\begin{alignat}{3}
\frac{c^2}{\gamma_\ms{a}}&=c^2\frac{d\tau_\ms{a}}{dt}
\\\nnm
&=c^2-\frac{v_\ms{a}^2}{2}+U-\frac{4}{c^2}v_\ms a^iU_i+\frac{O(v^2,U)^2}{c^2}+O(c^{-4}),
\end{alignat}
where the potentials $U$ and $U_i$ are evaluated at $z_\ms{a}$.  For our leading-order treatment of the current-quadrupole effects, which we can treat as a linear perturbation of the Newtonian point-mass dynamics, we will need to keep at $O(c^{-2})$ only the term linear in the vector potential $U_i$, while we can drop the other 1PN terms $\sim(v^4,\,v^2U,\,U^2)/c^2$.  The $v^iU_i$ term contributes to the current-quadrupole dynamics at leading order, even though it contributes to the point-mass dynamics only at next-to-leading (1PN) order, like the $O(v^2,U)^2$ terms.

Calculating the Weyl tensor of the 1PN metric (\ref{1PNmet}), one finds that its magnetic part $\mc B^1_{\mu\nu}$ with respect to $u_1^\mu$ at $z_1$ is given in the coordinate basis by
\begin{alignat}{3}\label{mcB}
  \mc{B}^1_{ij} & =\frac{2}{c^2} \epsilon^k{}_{l(i}\partial_{j)}\partial_{k}(Uv_1^l-U^l)+ O(c^{-4}),\nnm\\\nnm
  \mc{B}^1_{0i} & = -\frac{2}{c^2}v_1^j \epsilon^k{}_{l(i}\partial_{j)}\partial_{k}U^l + O(c^{-4}),\\
  \mc{B}^1_{00} & =  O(c^{-4}),
\end{alignat}
where the derivatives of the potentials are evaluated at $z_1$.  

Given that the current-quadrupole $S_1^{\mu\nu}$ is symmetric (and trace-free) and orthogonal to $u_1^\mu$, one can solve for the temporal components $S_1^{00}$ and $S_1^{0i}$ in terms of the purely spatial components $S_1^{ij}$ (and the components of $u_1^\mu$).  With $S_1^{ij}=O(c^{0})$, one finds $S_1^{00}=O(c^{-2})=S_1^{0i}$, and thus that $S_1^{ij}+O(c^{-2})$ is an STF spatial tensor, and that $S_1^{\mu\nu}\mc B^1_{\mu\nu}=S_1^{ij}\mc B^1_{ij}+O(c^{-4})$, given (\ref{mcB}).

Putting everything together, inserting spatial integrals and delta functions in $\mc S_1$ and $\mc S_2$ to evaluate the potentials and their derivatives along the worldlines, and using the notation $\doe_{jk}=\doe_j\doe_k$, our total binary action (\ref{binaryaction}) becomes
\begin{widetext}
\begin{alignat}{3}\label{Stotal}
\mc S= \int dt \, d^3 \bs{x} \, \Bigg \{ &\delta^3(\bs{x}-\bs{z}_1) \, \bigg [ m_1 \Big ( -c^2+\frac{v_1^2}{2}+U-\frac{4}{c^2}v_1^iU_i \Big)+\frac{4}{3c^2}S_1^{ij} \epsilon^k{}_{li} \partial_{jk}(Uv_1^l-U^l)+\mc L_1^\mr{int} \bigg ] 
\nnm\\
  +\,& \delta^3(\bs{x}-\bs{z}_2) \,  \bigg [ m_2 \Big( -c^2+\frac{v_2^2}{2}+U-\frac{4}{c^2}v_2^iU_i \Big) \bigg]+\frac{O(v^2,U)^2}{c^2}+ O(c^{-4}) \Bigg \}+\mc S_g,
\end{alignat}
where $\mc S_g$ is given by (\ref{Sg1PN}), noting that the internal Lagrangian $\mc L_1^\mr{int}$ depends only on the internal degrees of freedom $\psi_1$, not on the worldlines or potentials.

Varying the action with respect to the potentials $U$ and $U_i$ yields the field equations
\begin{alignat}{3}\label{UFES}
  \bs{\nabla}^2 U & =-4 \pi G \bigg [ \Big(m_1+\frac{4}{3c^2}S_1^{ij}v_1^l \epsilon^k{}_{li} \partial_{jk}\Big) \delta^3(\bs{x}-\bs{z}_1) + m_2 \delta^3(\bs{x}-\bs{z}_2)+\frac{O(v^2,U)^2}{c^2}  \bigg ]+\frac{\ddot{U}}{c^2}+ O(c^{-4}), 
  \\\label{UiFES}
  \bs{\nabla}^2U^i & = -4 \pi G  \bigg [ \Big(m_1v_1^i+\frac{1}{3}\epsilon^i{}_j{}^kS_1^{jl}\partial_{kl}\Big)\delta^3(\bs{x}-\bs{z}_1)+m_2v_2^i\delta^3(\bs{x}-\bs{z}_2) \bigg ]+ O(c^{-2}).
\end{alignat}  
Comparing these with (\ref{UFE}) and (\ref{UiFE}), we see that the forms match, and that we can read off the components $T^{00}+T^{ii}/c^2$ and $T^{0i}$ of the effective stress-energy tensor for our system; they are the quantities in square brackets in (\ref{UFES}) and (\ref{UiFES}) respectively.  The solutions which vanish at infinity are
\begin{alignat}{3}
  U(t,\bs{x}) & =\frac{Gm_1}{|\bs{x}-\bs{z}_1(t)|}+\frac{Gm_2}{|\bs{x}-\bs{z}_2(t)|}+\frac{4G}{3c^2}S_1^{ij}(t)v_1^l(t) \epsilon^k{}_{li} \partial_{jk}\frac{1}{|\bs{x}-\bs{z}_1(t)|}+ [\text{1PN p.m.}] + O(c^{-4}), 
  \label{Usol} 
  \\
  U^i(t,\bs{x}) & =\frac{Gm_1v_1^i(t)}{|\bs{x}-\bs{z}_1(t)|}+\frac{Gm_2v_2^i(t)}{|\bs{x}-\bs{z}_2(t)|}+\frac{G}{3}\epsilon^i{}_j{}^kS_1^{jl}(t)\partial_{kl}\frac{1}{|\bs{x}-\bs{z}_1(t)|}+ O(c^{-2}),
  \label{Uisol}
\end{alignat}
\end{widetext}
where we write [1PN p.m.]\ for the $O(c^{-2})$ point-mass terms that come from the $O(v^2,U)^2$ and $\ddot U$ terms in (\ref{UFES}), which we will not need.  Note that the $\ddot U$ term contributes $S_1^{ij}$ terms only at $O(c^{-4})$.\footnote{\label{foot:norm}We can now connect our normalization for $S^{ij}$, with the factor of 2/3 in (\ref{SEB}), to the definitions of the body's Blanchet-Damour multipole moments \cite{Blanchet:1989ki}.  From Eqs.~(3.8)--(3.10) of \cite{Vines:2010ca}, with $\Phi=-U$ and $\zeta_i=-4U_i$ (and restoring factors of $G$ set to 1 in \cite{Vines:2010ca}), dropping the external tidal terms and the $O(c^{-2})$ terms in $U$, the potentials generated some body are given in terms of its multipoles about some point $r=0$ by
\begin{alignat}{3}
G^{-1}U&=\frac{M}{r}-M^i\doe_i\frac{1}{r}+\frac{1}{2}M^{ij}\doe_{ij}\frac{1}{r}-\frac{1}{6}M^{ijk}\doe_{ijk}\frac{1}{r}+\ldots
\nnm\\
G^{-1}U^i&=\frac{\dot M^i}{r}-\frac{1}{2}\Big(\dot M^{ij}-\epsilon^{ij}{}_{k}S^k+\tfrac{1}{6}\delta^{ij}\mu\Big)\doe_j\frac{1}{r}
\\\nnm
&\quad+\frac{1}{6}\Big(\dot M^{ijk}-2\epsilon^{ij}{}_lS^{kl}+\tfrac{9}{20}\delta^{ij}\mu^k\Big)\doe_{jk}\frac{1}{r}+\ldots
\end{alignat}
where $\mu^{\ldots}$ are ``gauge moments.''  We see that the normalization of the $S^{ij}$ term in $U^i$ matches that in (\ref{Uisol}).
}

Having solved the field equations for the potentials $U$ and $U_i$, we can now insert these solutions into the action to find the reduced Fokker action depending only on the worldlines (and the internal degrees of freedom).  We encounter several divergent self-field terms, which we can simply drop; they are independent of the worldlines and thus would not affect the orbital equations of motion.   It is important that we use the total action (\ref{Stotal}), including $\mc S_g$; note the separate contributions
\begin{alignat}{3}
  \mc S_1+\mc S_2&=\int dt  \bigg[ \, \frac{1}{2}m_1v_1^2 + \frac{1}{2}m_2v_2^2+\frac{2Gm_1m_2}{|\bs{z}_1-\bs{z}_2|}
 \\\nnm 
&\quad  +\frac{8Gm_2}{3c^2}\, S_1^{ij}(v_1^l-v_2^l) \, \epsilon_i{}^k{}_l\, \partial^1_{jk}\frac{1}{|\bs{z}_1-\bs{z}_2|} +\mc L^{\mr{int}}_1 \bigg],
\end{alignat}
and
\begin{alignat}{3}
  \mc S_g&=\int dt  \bigg[ -\frac{Gm_1m_2}{|\bs{z}_1-\bs{z}_2|}
  \\\nnm
&\qquad\qquad  -\frac{4Gm_2}{3c^2}S^{ij}_1(v_1^l-v_2^l)\epsilon_i{}^k{}_l\partial^1_{jk}\frac{1}{|\bs{z}_1-\bs{z}_2|} \bigg],
\end{alignat}
where $\doe^1_i=\doe/\doe z_1^i$, and where we henceforth neglect to note the 1PN point-mass and $O(c^{-4})$ corrections.  Note that the effect of adding $\mc S_g$ is to halve all the potential terms, which were in a sense double-counted in $\mc S_1+\mc S_2$.
Finally, the total action (\ref{binaryaction}) becomes $\mc S=\int dt \, \mc L$ where 
\begin{alignat}{3}\label{Lgenframe}
\mc L&=\frac{1}{2}m_1v_1^2 + \frac{1}{2}m_2v_2^2+\frac{Gm_1m_2}{|\bs{z}_1-\bs{z}_2|}
\\\nnm
&\quad+\frac{4Gm_2}{3c^2}S^{ij}_1(v_1^l-v_2^l)\epsilon_i{}^k{}_l\partial^1_{jk}\frac{1}{|\bs{z}_1-\bs{z}_2|} +\mc L_1^{\text{int}}.
\end{alignat}
One can confirm that the $S_1^{ij}$ contributions to the orbital equations of motion obtained from varying the action $\int dt\,\mc L$ (\ref{Lgenframe}) with respect to $\bs z_1$ and $\bs z_2$ match those derived in \cite{Racine:2004hs}, as given in Eq.~(1) of \cite{PhysRevD.96.129904}.

The Lagrangian can be simplified further by specializing to the binary's center-of-mass frame.  We can fix the center of mass to be at rest at the origin by setting to zero the system's mass-dipole, which at the considered order yields the Newtonian relation $m_1\bs{z}_1+m_2\bs{z}_2=0$.\footnote{One can verify that there is no $c^{-2}S^{ij}$ contribution to the mass-dipole, either from Sec.~IV of \cite{Vines:2010ca} e.g., or by computing the mass dipole (at $t=0$) as the Noether charge of the boost symmetry of the Lagrangian (\ref{Lgenframe}).}  We thereby obtain a reduced Lagrangian given in terms of the relative position $\bs{r}=\bs{z}_1-\bs{z}_2$ and velocity $\bs{v}=\bs{v}_1-\bs{v}_2$,
\begin{equation}\label{Lcm}
\mc L=  \frac{\mu v^2}{2}+\frac{G \mu M}{r}+\frac{4Gm_2}{3c^2}S^{ij}_1v^l\epsilon_i{}^k{}_l \partial_{jk}\frac{1}{r}+\mc L_1^{\text{int}},
\end{equation}
where $r=|\bs r|$ and $\doe_i=\doe/\doe r^i$, and where we define the total mass $M=m_1+m_2$ and the reduced mass $\mu=m_1m_2/M$. 

The Lagrangian (\ref{Lcm}) determines the binary's orbital equation of motion, as the Euler-Lagrange equation for $\bs r(t)$.   The $S_1^{ij}$ term appears as a linear perturbation of the Newtonian (Keplerian) Lagrangian, in which $S_1^{ij}(t)$ is a yet unconstrained function of time which would be determined by the body's internal structure and dynamics.

\subsection{Adiabatic tidal current-quadrupole}\label{sec:adiabatic}

In the adiabatic approximation, assuming that body 1's internal dynamical timescales are small compared to the orbital timescale, it will develop a current-quadrupole proportional to the instantaneous gravitomagnetic tidal field, $S_1^{\mu\nu}\propto\mc B_1^{\mu\nu}$, or equivalently $S_1^{ij}\propto\mc B_1^{ij}$ at leading order \cite{Damour:1991yw,Favata:2005da,Binnington:2009bb,Damour:2009vw,Bini:2012gu}.  With this relation, $S_1^{ij}$ is determined by the orbital degrees of freedom, and we can obtain an action for the orbital dynamics depending only on $\bs r$ and $\bs v$.  The adiabatic relation $S_1^{ij}\propto\mc B_1^{ij}$ can be obtained from the general binary action (\ref{Lcm}) with a simple ansatz for the internal Lagrangain $\mc L_1^\mr{int}$.

Let us first note that, given the solutions (\ref{Usol})--(\ref{Uisol}) for the metric potentials in our binary, (the regular part of) the general expression (\ref{mcB}) for $\mc B_{ij}^1$ evaluates to
\begin{alignat}{3}\label{mcB1}
  \mc{B}^1_{ij}&= \frac{2Gm_2}{c^2} v^l \, \epsilon^k{}_{l(i}\partial_{j)k}\frac{1}{r}
  \\\nnm
  &=\frac{6Gm_2}{c^2r^3}v^l \, \epsilon^k{}_{l(i}n_{j)}n_k ,
  \label{eq17}
\end{alignat}
where $n_i\equiv r_i/r$.  We see that the Lagrangian (\ref{Lcm}) can be written as
\begin{equation}\label{LSB}
\mc L=  \frac{\mu v^2}{2}+\frac{G \mu M}{r}+\frac{2}{3}S_1^{ij}\mc B^1_{ij}+\mc L_1^{\text{int}}.
\end{equation}
To obtain an adiabatic current-quadrupole, we can treat the components $S_1^{ij}(t)$ themselves as the body's internal degrees of freedom $\psi_1(t)$, to be varied along with $\bs r(t)$ in the action, and posit that $\mc L_1^\mr{int}$ is quadratic in $S_1^{ij}$.  Choosing the coefficients so that $\sigma_1$ here (subscript 1 for body 1) matches the quadrupolar gravitomagnetic TLN $\sigma_1^{(2)}$ defined in \cite{Bini:2012gu}, we will obtain
\be\label{adrel}
S_1^{ij}=2c^2 \sigma_1 \mc{B}_1^{ij}
\ee 
from the Euler-Lagrange equation for $S_1^{ij}$ if
\be\label{L1int}
\mc L_1^{\text{int}}=-\frac{1}{6c^2\sigma_1}S^1_{ij}S_1^{ij}.
\ee
Inserting the solution (\ref{adrel}) into (\ref{LSB}) with (\ref{L1int}), we obtain a reduced Fokker Lagrangian for the orbital dynamics in the adiabatic approximation, with $\bs r(t)$ as the only remaining degree of freedom,
\begin{alignat}{3}\label{LBB}
\mc L&=\frac{\mu v^2}{2}+\frac{GM\mu}{r}+\frac{2}{3}c^2\sigma_1\mc B^1_{ij}\mc B_1^{ij}
\\\nnm
&=\frac{\mu v^2}{2}+\frac{GM\mu}{r}+\frac{12G^2m_2^2\sigma_1}{c^2r^6}\Big(v^2-(\bs v\cdot\bs n)^2\Big),
\end{alignat}
having used (\ref{mcB1}) in the second line.  One can confirm that this agrees with the leading gravitomagnetic contribution to the effective action for adiabatic tides derived in \cite{Bini:2012gu}; see in particular their Eq.~(4.8).

Since (\ref{LBB}) is rotation-invariant, the motion is confined to a plane.  It is convenient to use polar coordinates $(r,\phi)$ in the orbital plane, so that (\ref{LBB}) becomes
\be
\mc L=\frac{\mu}{2}(\dot r^2+r^2\dot\phi^2)+\frac{GM\mu}{r}+\frac{12G^2m_2^2\sigma_1\dot\phi^2}{c^2r^4}.
\ee
The resultant Euler-Lagrange equations for $r$ and $\phi$ read
\begin{equation}\label{ELr}
  \mu\ddot{r}= \mu\dot{\phi}^2r-\frac{GM\mu}{r^2}-\frac{48 G^2m_2^2\sigma_1 \dot{\phi}^2}{c^2r^5},
\end{equation}
\begin{equation}\label{ELphi}
 \frac{d}{dt}\left( \mu r^2 \dot{\phi}+ \frac{24G^2m_2^2\sigma_1\dot\phi}{c^2r^4}  \right)=0.
\end{equation}
The conserved energy resulting from time-translation invariance via $E=v^i(\doe\mc L/\doe v^i)-\mc L$ is
\begin{equation}\label{genE}
E=\frac{\mu}{2}(\dot r^2+r^2\dot\phi^2)-\frac{GM\mu}{r}+\frac{12G^2m_2^2\sigma_1\dot\phi^2}{c^2r^4}.
\end{equation}

\subsubsection*{Circular motion}

Specializing to circular orbits, for which $r$ is constant, it follows from (\ref{ELphi}) that $\dot{\phi}\equiv\omega$ is constant.  We can then solve for the orbital radius $r$ as a function of the orbital angular frequency $\omega$ by setting $\ddot r=0$ in (\ref{ELr}).  We find, to linear order in $\sigma_1$,
\be\label{romega}
r=\left(\frac{GM}{\omega^2}\right)^{1/3}\left(1+\frac{16m_2\sigma_1\omega^4}{m_1Mc^2}\right).
\ee
Using this in (\ref{genE}) yields the gauge-invariant expression for the energy as a function of the frequency,
\begin{alignat}{3}\label{Eomega}
E&=-\frac{\mu}{2}(GM\omega)^{2/3}\left(1-\frac{88m_2\sigma_1\omega^4}{m_1Mc^2}\right)
\nnm\\
&=-\frac{\mu c^2}{2}x\Big(1-88\nu X_1^3\Sigma_1 x^6\Big).
\end{alignat}
For the second line, we define the mass ratios
\be
X_1=\frac{m_1}{M},\quad X_2=\frac{m_2}{M},\quad \nu=X_1X_2=\frac{\mu}{M},
\ee
the dimensionless frequency parameter $x\sim v^2/c^2$ which counts PN orders,
\be\label{xomega}
x=\frac{(GM\omega)^{2/3}}{c^2},
\ee
and a dimensionless version $\Sigma_1$ of body 1's quadrupolar gravitomagnetic TLN $\sigma_1$,
\be\label{Sigma1}
\Sigma_1=\frac{G\sigma_1}{(Gm_1/c^2)^5}=\frac{k_1^\mr{2,mag}}{2}\left(\frac{\mc R_1c^2}{Gm_1}\right)^4,
\ee
where the stellar radius $\mc R_1$ and the dimensionless constant $k_2^\mr{mag}\to k_1^\mr{2,mag}$ for body 1 are as in (\ref{sigmaR}).

\section{Gravitational radiation}\label{sec:rad}

\subsection{System multipole moments}

The far-zone (post-Minkowskian) gravitational-wave field is determined in the PN approximation by the multipole moments of the entire binary system, which are encoded in the near-zone PN gravitational field \cite{Blanchet:2013haa}.  We will denote the system's mass-multipoles by $I_L$ and its current-multipoles by $J_L$, using the multi-index notation $L=i_1\ldots i_\ell$.  For our purposes, at the order of the 1PN field equations, these moments agree with both the source moments $I_L$ and $J_L$ and the canonical moments $M_L$ and $S_L$ defined in \cite{Blanchet:2013haa}.

The system multipoles can be computed in two different ways.  Firstly, if one has the components of the system's (effective) stress-energy tensor $T^{\mu\nu}$, then the moments (about $x^i=0$, which we assume has been fixed to the system's center of mass) are given at 1PN order by the integrals \cite{Blanchet:2013haa}
\begin{alignat}{3}\label{IJint}
  I^L&=\int d^3\bs x\, \bigg[\bigg(T^{00}+\frac{T^{jj}}{c^2}\Big)x^{<L>}
 +\frac{\ddot{T}^{00}}{2(2\ell+3)c^2}x^{jj<L>}
\nnm\\\nnm
&\qquad\qquad\quad-\frac{4(2\ell+1)\dot{T}^{0j}}{(\ell+1)(2\ell+3)c^2}x^{<jL>} \bigg]+O(c^{-4}) ,
\\
  J^L&=\int  d^3\bs x \;T^{0k}\epsilon^{jk<i_\ell}x^{L-1>j}+O(c^{-2}) ,
\end{alignat}
where $x^L=x^{i_1}\ldots x^{i_\ell}$, etc.  For our $m_1$-$S_1^{ij}$-$m_2$ system from Sec.~\ref{sec:cons}, the components $T^{00}+T^{ii}/c^2$ and $T^{0i}$ are given respectively by the quantities in square brackets in the first and second lines of (\ref{Stotal}).  Inserting these into (\ref{IJint}) for $\ell=2,3$, integrating, and using (only) the leading-order center-of-mass-frame relations $m_1\bs z_1=-m_2\bs z_2=\mu\bs r$ (not yet specializing to adiabatic tides or to circular orbits), we find the system's quadrupoles and octupoles:
\begin{alignat}{3}\label{IJ23}
  I^{ij}&=\mu  r^{<ij>}+\tr{[1PN p.m.]}
  \nnm\\
  &\quad+\frac{8X_2}{9c^2}\epsilon^{kl<i}\Big(2S_1^{j>l}v^k-\dot{S}_1^{j>l}r^k\Big)+O(c^{-4}),
  \nnm\\
  J^{ij}&=-\delta\mu  \epsilon^{kl<i}r^{j>k}v^l+S_1^{ij}+O(c^{-2}),\phantom{\Big|}
  \nnm\\
  I^{ijk}&=-\delta\mu r^{<ijk>}+\tr{[1PN p.m.]}
 \\
  &\quad+\frac{2}{c^2}X_2^2\epsilon^{lm<i}\Big(3r^jS_1^{k>m}v^l
\nnm\\
&\qquad\qquad\qquad\;\;\,-v^jS_1^{k>m}r^l-r^j\dot S_1^{k>m}r^l\Big)+O(c^{-4}),
    \nnm\\\nnm
  J^{ijk}&=(1-3\nu)\mu\epsilon^{lm<i}r^{jk>l}v^m+\frac{8}{3}X_2S_1^{<ij}r^{k>}+O(c^{-2}),
\end{alignat}
with $r^{ij}=r^ir^j$, etc., and where we define the antisymmetric mass ratio
\be
\delta=\frac{m_1-m_2}{M}=X_1-X_2.
\ee
Note that the $\ddot T^{00}$ term in $I^L$ in (\ref{IJint}) contributes only to the omitted 1PN point-mass terms here.

In a second equivalent way to compute the system multipoles at relative 1PN order, one can start from the expression for the near-zone 1PN metric potentials (rather than starting from a stress-energy tensor), as discussed in Sec.~IV of \cite{Vines:2010ca}.  In Sec.~\ref{sec:modes}, we describe the implementation of that procedure to compute the same multipoles given in (\ref{IJ23}), while also including a mass-quadrupole $M_1^{ij}$ for body 1, working to relative 1PN order in the $M_1^{ij}$ terms.

\subsection{Energy flux and waveform phasing for adiabatic tides and quasi-circular orbits}\label{sec:flux}

The energy flux (or power) $\dot E$ carried away from the system by gravitational radiation is given in terms of the system multipoles, to relative 1PN order, by \cite{Blanchet:2013haa}
\begin{alignat}{3}  \label{genflux}
  \dot{E}&=-\frac{G}{5c^5}\langle{I}^{(3)}_{ij}{I}^{(3)}_{ij}\rangle
  \\\nnm
  &\quad-\frac{G}{9c^7} \Big[\frac{1}{21}\langle{I}^{(4)}_{ijk}{I}^{(4)}_{ijk}\rangle+\frac{16}{5}\langle{J}^{(3)}_{ij}{J}^{(3)}_{ij}\rangle\Big]+ O(c^{-8}) , 
\end{alignat}
where $F^{(n)}=d^n F/dt^n$, and where $\langle\ldots\rangle$ denotes a suitable time-average (which is trivial for the case of circular orbits).  The time-derivatives of the system multipoles are to be computed using the PN conservative dynamics, which we will take here to be the circular orbit with adiabatic gravitomagnetic tides.

In Cartesian coordinates $(X,Y,Z)$, for circular orbits in the $X$-$Y$-plane, we have 
\begin{alignat}{3}\label{lambda}
r^i(t)&=r(\cos\omega t,\sin\omega t,0)=rn^i(t),
\nnm\\
v^i(t)&=r\omega(-\sin\omega t,\cos\omega t,0)\equiv r\omega \lambda^i(t),
\nnm\\
e_Z^i&=\epsilon^{ijk}n^j\lambda^k,
\end{alignat}
 where $\lambda^i$ is the unit vector in the direction of the velocity and $e_Z^i$ is the unit vector in the $Z$-direction.  The time derivatives of the system multipoles are most easily calculated by first expressing them in terms of only $n^i(t)$, $\lambda^i(t)$, $e_Z^i$ and constants, and using 
\be\label{ndot}
\dot n^i=\omega\lambda^i,\qquad \dot\lambda^i=-\omega n^i,\qquad \dot e_Z^i=0.
\ee
Using the adiabatic relation (\ref{adrel}) with (\ref{mcB1}) to replace $S_1^{ij}$, we find the following Newtonian point-mass and leading-order gravitomagnetic tidal contributions (dropping others) to the system quadrupoles from (\ref{IJ23}),
\begin{alignat}{3}\label{IJ2}
  I^{ij}&=\mu r^2 n^{<ij>}+\frac{16GMX_2^2\omega^2\sigma_1}{3c^2r} (5n^{<ij>}+4\lambda^{<ij>}),
  \nnm\\
  J^{ij}&=\bigg(\!-\delta\mu r^3\omega+\frac{12GMX_2\omega\sigma_1}{r^2}\bigg)n^{<i}e_Z^{j>}.
\end{alignat}
Note that only these $\sigma_1$ terms will contribute to the flux (\ref{genflux}) at 1PN order; the $S_1^{ij}$ term in $I^{ijk}$ (or $J^{ijk}$) would contribute to $\dot E$ at $O(c^{-4})$.

It is then straightforward to thrice differentiate (\ref{IJ2}) and insert the results into (\ref{genflux}), using (\ref{romega}) to eliminate $r$ in favor of $\omega$.  Dropping all contributions except for the Newtonian point-mass and leading $\sigma_1$ terms, and using the definitions (\ref{xomega})--(\ref{Sigma1}), we find the flux to be
\be\label{Edot}
\dot E=-\frac{32 c^5}{5G}\nu^2 x^5\Big(1+\frac{2}{3}X_1^4(114 X_2-1)\Sigma_1x^6\Big).
\ee
In the stationary phase approximation, the phase $\psi$ of the frequency-domain GW signal is determined as a function of the orbital angular frequency $\omega$ by the flux $\dot E(\omega)$ and the conservative orbital energy $E(\omega)$ via \cite{Tichy:1999pv}
\be
\frac{d^2\psi}{d\omega^2}=\frac{2}{\dot E}\frac{dE}{d\omega}.
\ee
Substituting our results (\ref{Eomega}) and (\ref{Edot}), working again to linear order in the tidal perturbation, and twice integrating (dropping integration constants), we find
\be\label{psiomega}
\psi(\omega)=\frac{3}{128\nu x^{5/2}}\Big(1-\frac{20}{21}X_1^4(1038 X_2-1)\Sigma_1x^6\Big).
\ee
Our result (\ref{Edot}) for the flux matches that first derived in 
\cite{PhysRevD.96.129904} and later confirmed in \cite{Abdelsalhin:2018reg,Landry:2018bil}.  Our result (\ref{psiomega}) for the phase matches those found in \cite{PhysRevD.96.129904,Abdelsalhin:2018reg}.

\subsection{Waveform mode amplitudes for adiabatic tides and quasi-circular orbits}\label{sec:modes}

The far-zone post-Minkowskian gravitational wave field can also be expressed directly in terms of the system's (PN) multipole moments \cite{Blanchet:2013haa,Kidder:2007rt}.  We will follow here the conventions of Sec.~II of \cite{Kidder:2007rt}.  The polarization waveforms $h_+$ and $h_\times$ defined by Eqs.~(6) and (10) of \cite{Kidder:2007rt} are given by Eq.~(11) of \cite{Kidder:2007rt} as
\be
h_+-ih_\times=\sum_{\ell=2}^\infty\sum_{m=-\ell}^\ell h_{\ell m}(T_R)\,_{-2}Y^{\ell m}(\Theta,\Phi),
\ee
where $_{-s}Y^{\ell m}$ are the spin-weighted spherical harmonics as defined by Eqs.~(4)--(5) of \cite{Kidder:2007rt}, in the spherical radiative coordinate system $(T,R,\Theta,\Phi)$, with $T_R=T-R/c$ being the retarded time.  The mode amplitudes $h_{\ell m}$ are given by Eqs.~(13) and (19) of \cite{Kidder:2007rt} as
\be\label{hlmgen}
h_{\ell m}=\frac{8\pi G}{Rc^{\ell+2}}\sqrt{\frac{(\ell+2)(\ell+1)}{\ell(\ell-1)}}\bigg(
\mc U_L+\frac{2i\ell \mc V_L}{(\ell+1)c}
\bigg)\frac{(\mc Y_{\ell m}^L)^*}{(2\ell+1)!!},
\ee
where $\mc U_L$ and $\mc V_L$ are the system's radiative multipole moments, given to relative 1PN order by the $\ell$th time derivatives of the PN source/canonical multipoles $I_L(t)$ and $J_L(t)$ evaluated at the retarded time,
\begin{alignat}{3}\label{UV}
\mc U_L&=I^{(\ell)}_L(T_R)+O(c^{-3}),
\\\nnm 
\mc V_L&=J^{(\ell)}_L(T_R)+O(c^{-3}),
\end{alignat}
and where $(\mc Y_{\ell m}^L)^*$ are the complex conjugates of the STF spherical harmonics $\mc Y_{\ell m}^L$.  The STF harmonics are related to the usual scalar spherical harmonics $Y^{\ell m}={}_0Y^{\ell m}$ and the radial unit vector $N^i$ by
\begin{alignat}{3}
\mc Y_{\ell m}^L N^L&=Y^{\ell m}(\Theta,\Phi),
\\\nnm
N^i&=\sin\Theta(\cos\Phi\, e_X^i+\sin\Phi\, e_Y^i)+\cos\Theta\, e_Z^i,
\end{alignat}
where $(e_X^i, e_Y^i, e_Z^i)$ is the Cartesian orthonormal triad.  Defining the complex vector
\be
\zeta^i=e_X^i+ie_Y^i,
\ee
the STF harmonics $\mc Y_{\ell m}^L$ for $\ell=2,3$ and $m\ge0$ are given explicitly by
\begin{alignat}{3}\label{Y23}
\mc Y_{22}^{ij}&=\frac{1}{4}\sqrt{\frac{15}{2\pi}}\zeta^{<ij>},
\qquad&
\mc Y_{21}^{ij}&=-\frac{1}{2}\sqrt{\frac{15}{2\pi}}\zeta^{<i}e_Z^{j>},
\nnm\\
\mc Y_{20}^{ij}&=\frac{3}{4}\sqrt{\frac{5}{\pi}}e_Z^{<ij>},
\\\nnm
\mc Y_{33}^{ijk}&=-\frac{1}{8}\sqrt{\frac{35}{\pi}}\zeta^{<ijk>},
\qquad&
\mc Y_{32}^{ijk}&=\frac{1}{4}\sqrt{\frac{105}{2\pi}}\zeta^{<ij}e_Z^{k>},
\\\nnm
\mc Y_{31}^{ijk}&=-\frac{5}{8}\sqrt{\frac{21}{\pi}}\zeta^{<i}e_Z^{jk>},
\qquad&
\mc Y_{30}^{ijk}&=\frac{5}{4}\sqrt{\frac{7}{\pi}}e_Z^{<ijk>}.
\end{alignat}
For circular orbits in the $X$-$Y$-plane, as in (\ref{lambda}) with $\omega t=\phi$ (where we can identify the Cartesian frame there with that used here), it is useful to note that 
\be\label{nzeta}
n^i+i\lambda^i=e^{-i\phi}\zeta^i.
\ee
Finally, note that the mode amplitudes with $m<0$ can be found from those $m>0$ via $h_{\ell,-m}=(-1)^\ell (h_{\ell m})^*$, as in Eq.~(78) of \cite{Kidder:2007rt}.

Equations (\ref{hlmgen})--(\ref{nzeta}) allow an explicit computation of the amplitudes $h_{\ell m}$ for $\ell=2,3$, to relative 1PN order, given expressions for the PN source quadrupoles $I_{ij}$ and $J_{ij}$ and octupoles $I_{ijk}$ and $J_{ijk}$.  For our $m_1$-$S_1^{ij}$-$m_2$ system, these are given above in (\ref{IJ23}), for general orbits and arbitrary $S_1^{ij}$.  We gave the specializations of $I_{ij}$ and $J_{ij}$ to adiabatic tidal $S_1^{ij}$ and to circular orbits in (\ref{IJ2}), and the specializations of $I_{ijk}$ and $J_{ijk}$ can be found analogously.

For comparison with the contributions from gravitomagnetic tides $\propto\sigma_1\propto\Sigma_1$, we will also include here the contributions to $h_{2m}$ and $h_{3m}$ from the leading adiabatic gravitoelectric tides, proportional to the quadrupolar gravitoelectric TLN $\lambda_1$ for body 1, with an adiabatic tidal mass quadrupole $M_1^{ij}=-\lambda_1\mc E^{ij}$, working to relative 1PN order.  For this, we borrow several results from \cite{Vines:2010ca,Vines:2011ud}, and we compute for the first time the relative-1PN gravitoelectric tidal contributions to $I_{ijk}$ and $J_{ijk}$.  We will also restore here the 1PN point-mass terms which we have thus far dropped.

To compute all of those contributions to the system quadrupoles and octupoles, also reproducing the $S_1^{ij}$ contributions given above in (\ref{IJ23}), one can use Eqs.~(4.5) and (4.6) of \cite{Vines:2010ca}, yielding the system multipoles $I^L\equiv M_\mr{sys}^L$ and $J^L\equiv S^L_\mr{sys}$ (for $\ell=2,3$), via the intermediate moments $Z^{iL}_\mr{sys}$ (for $\ell=2,3,4$) and $\mu^L_\mr{sys}$ (for $\ell=2,3$).  This gives the system multipoles in terms of the global-frame body multipoles $M_{\mr g,A}^L$ and $Z^{iL}_{\mr g,A}$ for each body $A=1,2$.  The latter are given in Eqs.~(B4) and (B5) of \cite{Vines:2010ca}, which include the mass quadrupole $Q_{ij}$ for one of the bodies.  In using those equations from \cite{Vines:2010ca}, as well as others from \cite{Vines:2011ud} below, one must note that those references took body 2 to be the one with a mass-quadrupole, and they defined the relative position $z^i=z_2^i-z_1^i$, as opposed to our definition $r^i=z_1^i-z_2^i$ here.  Thus, to translate from \cite{Vines:2010ca,Vines:2011ud} to our conventions, one should simply exchange $1\leftrightarrow 2$ everywhere in \cite{Vines:2010ca,Vines:2011ud} (\emph{without} flipping the signs of $z^i\to r^i$, $n^i$, $v^i$, etc.).  The global-frame body multipoles we need to insert into Eqs.~(4.5) and (4.6) of \cite{Vines:2010ca} are thus given by (B4) and (B5) of \cite{Vines:2010ca} under $1\leftrightarrow2$, where then $Q^{ij}\equiv M_1^{ij}$---plus contributions from body 1's current-quadrupole $S_1^{ij}$.  These can be found either from Eqs.~(B2) and (B3) of \cite{Vines:2010ca}, or by comparing Eqs.~(4.1) of \cite{Vines:2010ca} to our (\ref{Usol})--(\ref{Uisol}) above (with $\Phi_\mr{g}=-U$ and $\zeta^i_\mr{g}=-4U^i$).  One finds that the necessary additions are given by
\begin{alignat}{3}
M_{\mr g,1}^{ij}&\to M_{\mr g,1}^{ij}+\frac{8}{3c^2}v_1^k\epsilon^{kl(i}S_1^{j)l},
\nnm\\
Z_{\mr g,1}^{ijk}&\to Z_{\mr g,1}^{ijk}+\frac{8}{3}\epsilon^{il(j}S_1^{k)l},
\end{alignat}
with all the other moments in (B4)--(B5) of \cite{Vines:2010ca} (under $1\leftrightarrow2$) unchanged, and with those being the only nonzero global-frame body moments.  These involve also the spin $S^i=S_2^i\to S_1^i$ (which cannot be dropped when treating mass-quadrupoles at 1PN order in general, as discussed in \cite{Vines:2010ca}, but which can be consistently dropped for adiabatic tidal mass-quadrupoles).  Finally, one must note that the Blanchet-Damour mass monopole $M_2$ $(\to M_1)$ appearing in the second line of Eq.~(B4) of \cite{Vines:2010ca} is not a constant.  It is given by Eq.~(3.30) in terms of ${}^{\tr{\tiny n}\!}M_2$ which is a constant ($\to m_1$ here), $U_Q$ given by Eq.~(2.28), and $E_2^\mr{int}$ [which is given by Eq.~(6.3) in the adiabatic approximation]; all Eqs.\ from \cite{Vines:2010ca}.  The other monopole, of the body without higher multipoles, is a constant; the $M_1$ there becomes $m_2$ here.  Note also $\chi_{2,1}\to X_{1,2}=m_{1,2}/M$.

One then has the system multipoles $I^L=M_\mr{sys}^L$ and $J^L=S_\mr{sys}^L$ expressed in terms of $m_1$ and $m_2$, and $z_1^i$, $z_2^i$, $S_1^i$, $M_1^{ij}$, $S_1^{ij}$, $E^\mr{int}_1$, and their time derivatives.  To express $z_1^i$ and $z_2^i$ in terms of $r^i=z_1^i-z_2^i$, one must use the $1\leftrightarrow2$ translation of Eqs.~(5.2)--(5.4) of \cite{Vines:2010ca}, which result from setting to zero the system's mass dipole $M_\mr{sys}^i$, given by Eq.~(4.8) of \cite{Vines:2010ca}.  One can confirm that there are no $S_1^{ij}$ contributions to $M_\mr{sys}^i$ at 1PN order by also computing it via the procedure outlined in the previous paragraph.  Next, to specialize to  adiabatic tidal multipoles, one uses Eqs.~(6.3) and (6.6) of \cite{Vines:2010ca} with $1\leftrightarrow2$ for $Q^{ij}\to M_1^{ij}$ and $E^\mr{int}_2\to E^\mr{int}_1$ and our (\ref{mcB1}) and (\ref{adrel}) above for $S_1^{ij}$.  One can then also consistently drop all contributions from the spin $S_1^i$.

At this point, one has the system multipoles in terms of only $m_1$, $m_2$, the quadrupolar gravitoelectric TLN $\lambda\to\lambda_1$ for body 1, its quadrupolar gravitomagnetic TLN $\sigma_1$, and the relative position $r^i$ and its time derivatives.  One can next specialize to the case of circular orbits, as in (\ref{lambda}) above.  Here one requires the radius-frequency relation $r(\omega)$ resulting from the conservative orbital equations of motion.  The point-mass and $\lambda\to\lambda_1$ contributions to $r(\omega)$ are given to 1PN order by Eq.~(2.9)--(2.10) of \cite{Vines:2011ud} under $1\leftrightarrow2$, and to this we add the $\sigma_1$ contribution in our (\ref{romega}) above.  After this, one has the system multipoles in terms of $m_1$, $m_2$, $\lambda_1$, $\sigma_1$, the orbital angular frequency $\omega$, and the unit vectors $n^i$ and $\lambda^i$ as in (\ref{lambda}) and their time derivatives which are easily computed via (\ref{ndot}).  As we did for $\sigma_1$ in (\ref{Sigma1}), we define the dimensionless version of $\lambda_1$,
\be\label{Lambda1}
\Lambda_1=\frac{G\lambda_1}{(Gm_1/c^2)^5}=\frac{2}{3}k_1^\mr{2,el}\left(\frac{\mc R_1c^2}{Gm_1}\right)^5,
\ee
where the stellar radius $\mc R_1$ and the dimensionless constant $k_2^\mr{el}\to k_1^{2,\mr{el}}$ for body 1 are as in (\ref{lambdak}).

The resultant expressions for $I_{ij}$, $J_{ij}$, $I_{ijk}$, and $J_{ijk}$ can be inserted into (\ref{hlmgen})--(\ref{UV}), computing the time derivatives with (\ref{ndot}), and computing the various STF contractions of the unit vectors, to find the mode amplitudes $h_{2m}$ and $h_{3m}$.

Finally, in the following expressions for the modes, we let body 2 also have adiabatic tidal quadrupoles $M_2^{ij}$ and $S_2^{ij}$, with dimensionless TLNs $\Lambda_2$ and $\Sigma_2$.  The extra contributions can be found simply by exchanging the bodies' identities---being careful to note that this involves sign-flips for the contributions to the odd-$m$ modes.  One can see that the tidal contributions to the odd-$m$ modes must be antisymmetric under exchange of the bodies' identities, e.g.\ for the $S_{ij}$ contributions, as follows.  Firstly, $r^i=rn^i=z_1^i-z_2^i$ and $v^i=v_1^i-v_2^i$ flip sign under the exchange, and thus also $S_1^{ij}\leftrightarrow -S_2^{ij}$ since $\mc B_1^{ij}\leftrightarrow -\mc B_2^{ij}$ due to two powers of $n^i$ and one power of $v^i$ in (\ref{mcB}).  One then sees from (\ref{IJ23}) that the $S_{ij}$ contributions to $J_{ij}$ and $I_{ijk}$ flip signs while those to $I_{ij}$ and $J_{ijk}$ do not; a similar analysis leads to the same conclusions for the $M_{ij}$ contributions.  Finally, we note that, exclusively, $I_{ij}$ contributes to $h_{22}$, $J_{ij}$ to $h_{21}$, $I_{ijk}$ to $h_{33}$ and $h_{31}$, and $J_{ijk}$ to $h_{32}$.

Using the dimensionless frequency parameter $x$ from (\ref{xomega}), our final results for the mode amplitudes, through relative 1PN order in the point-mass and tidal terms, are
\begin{widetext}
\begin{alignat}{11}\label{modeseq}
h_{22}&=-\frac{GM}{Rc^2}8\nu x e^{-2i\phi}\sqrt{\frac{\pi}{5}}\Bigg\{\Big(1
-\frac{107-55\nu}{42}x+O(x^{3/2})\Big)
\\\nnm
&\qquad\qquad\qquad\qquad\qquad+\bigg[\Lambda_1 x^5 X_1^4\Big(3(1+2X_2)+\frac{63 - 15 X_2 - 205 X_2^2 - 45 X_2^3}{14}x+O(x^{3/2})\Big)
\\\nnm
&\qquad\qquad\qquad\qquad\qquad\;\;+\Sigma_1 x^6 X_1^4\Big(\frac{112}{3}X_2+O(x)\Big)+(1\leftrightarrow2)\bigg]\Bigg\},
\\\nnm
h_{21}&=-\frac{GM}{Rc^2}\frac{8i}{3}\nu x^{3/2}e^{-i\phi}\sqrt{\frac{\pi}{5}}\Bigg\{\delta\Big(1+O(x)\Big)+\bigg[\Lambda_1x^5X_1^4\Big(9X_2\frac{1-4 X_2}{2}+O(x)\Big)
\\\nnm
&\qquad\qquad\qquad\qquad\qquad\qquad\qquad\qquad\qquad\,\;+\Sigma_1 x^5 X_1^4\Big(-12+O(x)\Big)-(1\leftrightarrow2)\bigg]\Bigg\},
\\\nnm
h_{33}&=\frac{GM}{Rc^2}3i\nu x^{3/2}e^{-3i\phi}\sqrt{\frac{6\pi}{7}}\Bigg\{\delta\Big(1-2(2-\nu)x+O(x^{3/2})\Big)
\\\nnm
&\qquad\qquad\qquad\qquad\qquad\quad\;+\bigg[\Lambda_1 x^5 X_1^4\Big(-18 X_2^2+3X_2\frac{-2 + 9 X_2 + 25 X_2^2 + 10 X_2^3}{2}x+O(x^{3/2})\Big)
\\\nnm
&\qquad\qquad\qquad\qquad\qquad\quad\;\;\;+\Sigma_1 x^6 X_1^4\Big(12X_2(4-9X_2)+O(x)\Big)-(1\leftrightarrow2)\bigg]\Bigg\},
\\\nnm
h_{32}&=-\frac{GM}{Rc^2}\frac{8}{3}\nu x^{2}e^{-2i\phi}\sqrt{\frac{\pi}{7}}\Bigg\{\Big(1-3\nu+O(x)\Big)+\bigg[\Lambda_1x^5X_1^4\Big(12X_2(1-2X_2+3X_2^2)+O(x)\Big)
\\\nnm
&\qquad\qquad\qquad\qquad\qquad\qquad\qquad\qquad\qquad\quad\;\;+\Sigma_1 x^5 X_1^4\Big(32 X_2+O(x)\Big)+(1\leftrightarrow2)\bigg]\Bigg\},
\\\nnm
h_{31}&=-\frac{GM}{Rc^2}\frac{i}{3}\nu x^{3/2}e^{-i\phi}\sqrt{\frac{2\pi}{35}}\Bigg\{\delta\Big(1-\frac{2}{3}(4+\nu)x+O(x^{3/2})\Big)
\\\nnm
&\qquad\qquad\qquad\qquad\qquad\quad\;\;+\bigg[\Lambda_1 x^5 X_1^4\Big(-18 X_2^2+X_2\frac{10 -133 X_2 +259 X_2^2 - 130 X_2^3}{2}x+O(x^{3/2})\Big)
\\\nnm
&\qquad\qquad\qquad\qquad\qquad\quad\;\;\;\;+\Sigma_1 x^6 X_1^4\Big(12X_2(4-17X_2)+O(x)\Big)-(1\leftrightarrow2)\bigg]\Bigg\},
\end{alignat}
\end{widetext}
while $h_{20}$ is zero up through the orders given here for $h_{22}$, and $h_{30}$ is zero up through the orders given here for $h_{32}$.

The point-mass terms match those given e.g.\ in \cite{Kidder:2007rt}.  The $\Lambda x^5$ terms in $h_{22}$, $h_{21}$, $h_{33}$ and $h_{31}$ and the $\Lambda x^6$ term in $h_{22}$ match those first computed in \cite{Damour:2012yf}.  The $\Lambda x^6$ terms in $h_{33}$ and $h_{31}$, the $\Lambda x^5$ term in $h_{32}$, and all of the $\Sigma$ terms have been computed for the first time here.

\section{Discussion}

We have considered the leading-order effects of the bodies' current-quadrupole moments $S_{ij}$ on the dynamics of a binary system according to GR, eventually specializing to the case of adiabatic tidal $S_{ij}$'s to find the leading gravitomagnetic tidal effects.  We showed how the conservative contributions to the dynamics from arbitrary $S_{ij}$'s can be efficiently computed from an effective action approach, starting directly from the general 1PN metric, which greatly streamlines the calculations compared to previous treatments.

We went on to calculate the leading gravitomagnetic tidal effects in the GWs emitted by the binary, reproducing recent results for the phase of the frequency-domain PN waveform, and deriving for the first time the leading gravitomagnetic tidal contributions to the amplitudes of the spherical harmonic modes of the waveform.  We gave the mode amplitudes $h_{\ell m}$ for $\ell=2$ and 3 including all adiabatic tidal effects through relative 1PN order, which also includes new contributions from gravitoelectric tides.

We see in (\ref{modeseq}) that, in $h_{21}$ and $h_{32}$, the gravitomagnetic $\Sigma$ terms contribute at the same leading PN orders as the gravitoelectric $\Lambda$ terms, namely $O(x^5)$ or 5PN order relative to the leading point-mass terms.  This is in contrast to the other modes (and the phase) in which the leading $\Sigma$ terms are suppressed by a factor of $x$ relative the $\Lambda$ terms, the former starting at 6PN order and the latter at 5PN order.  While $h_{21}$ vanishes for equal masses and equal TLNs, $h_{32}$ does not.  These results may prove useful in comparisons between PN waveforms and numerical simulations of inspiralling binary NSs, and in parameter estimation studies of GW signals.

\begin{acknowledgments}

We thank Tiziano Abdelsalhin, Alessandra Buonanno, Marc Favata, Leonardo Gualtieri, Tanja Hinderer, Sylvain Marsat, Paolo Pani, and Jan Steinhoff for useful discussions, and we are very grateful to Tiziano Abdelsalhin for pointing out errors in an earlier version of our calculations.

\end{acknowledgments}


%

\end{document}